\documentclass[a4paper,10pt,orivec]{llncs}

\usepackage[utf8]{inputenc}
\usepackage{amsmath}
\usepackage[english]{babel}
\usepackage{graphicx}
\usepackage{multicol}
\usepackage{subfigure}
\title{Community-detection cellular automata with local and long-range connectivity}

\author{Franco Bagnoli\thanks{franco.bagnoli@unifi.it}\inst{1,2}\and
Andrea Guazzini\thanks{andrea.guazzini@iit.cnr.it}\inst{3}\and
Emanuele Massaro\thanks{emanuele.massaro@unifi.it}\inst{1}
}
\institute{
 Dept. Energy and CSDC, Universit\`a di Firenze, \\via S. Marta, 3 50139 Firenze, Italy. \\
 \and
 Also INFN, sez. Firenze.
  \and
  Dept. Psychology and CSDC, Università di Firenze, and  Institute for Informatics and Telematics (IIT), 
           National Research Council (CNR), \\
           via G. Moruzzi, 1 56124 Pisa, Italy     
}
\date{5/2/2011}

\begin{document}
 \maketitle
 
 \thispagestyle{empty}

\paragraph{Keywords:} Graph theory, Community, Clustering

\begin{abstract}
 We explore a community-detection cellular automata algorithm inspired by human heuristics, based on information diffusion  and a non-linear processing phase with a dynamics inspired by human heuristics. The main point of the methods is that of furnishing different ``views'' of the clustering levels from an individual point of view.  We apply the method to networks with local connectivity and long-range rewiring. 
\end{abstract}

\section{Introduction}
Detecting communities is a task of great importance in many disciplines, namely sociology, biology and computer science~\cite{Waaserman,Scott,Mendes,Strogatz,Albert},  where systems are often represented as graphs. Community detection is linked to clustering of data: many clustering methods establish links among representative points that are nearer than a given threshold, and then proceed in identifying communities on the resulting graphs~\cite{clustering1,domany}. 
Given a graph, a community is a group of vertices ``more linked'' than between the group and the rest of the graph. This is clearly a poor definition, and indeed, on a connected graph, there is not a clear distinction between a community and a rest of the graph. In general, there is a continuum of nested communities whose boundaries are somewhat arbitrary: the structure of communities can be seen as a hierarchical dendogram~\cite{Newman}. 

The problem of community detection is fundamental for social simulations: human decisions (for instance, cooperation vs. exploitation) often depends on the detection of the community in which one is embedded~\cite{Nowak}. 

Community detection relies on global  quantities like betweenness, centrality, etc.~\cite{Newman,communities} and the most algorithms require that the graph be completely known. This constraint is problematic for networks like the World Wide Web, which for all practical purposes is too large and too dynamic to ever be known fully. In  2005 Clauset~\cite{Clauset2005}  has introduced the concept of \emph{local modularity} because he wanted to find a local measure to remedy at the global measure of \emph{modularity} defined by Girvan and Newman in 2004\cite{Newman}.

For instance, let us suppose to be an internet user, who wants to know, at a certain time, which community belongs. For the user will be impossible to calculate the network betweenness (for example) of the World Wide Web: besides he could use an individual heuristic to discover the first neighbours node that he is connected determining in the base of his lack the best community for him.

At a superficial level, most of our information processing concerns the evaluation of probabilities. When faced with insufficient data or insufficient time for a rational processing, we humans have developed algorithms, denoted heuristics, that allows to take decisions in these situations. The modern approach to the study of cognitive heuristics defines them as those \textit{strategies that prevent one from finding out or discovering correct answers to problems that are assumed to be in the domain of probability theory}. 

Basically, the cognitive heuristics program proposed by Goldstein and Gigerenzer suggests to start from fundamental psychological mechanisms in order to design the models of heuristics~\cite{Gigerenzer2002}. 

Here we propose a new tool for detecting communities in complex networks using a local algorithm, applied as an (irregular) cellular automaton. In our previous work we shown an information dynamics algorithm, inspired by human heuristics, capable to discover communities in regular networks~\cite{Massaro11}.

In our approach an individual (node) is simply modelled as a memory and a set of connections to other individuals. The information about neighbouring nodes is propagated using a standard diffusion process, and elaborated locally using a non-linear competition process among the information. This process can be considered an implementation of the ``take the best'' heuristic~\cite{Gigerenzer1996}, which is simply the assumption that the most vivid or easily recallable information give an accurate estimate of the frequency of the related event in the population. 

The applicability of community detection algorithms to a network with local connectivity is rather problematic. Let us consider for illustration the small-world effect~\cite{SmallWorld}: starting from a regular network with pure local connectivity a small fraction of links is rewired to other sites. What is generally observed is that local quantities (like the clustering level) do not change until the fraction of rewiring is quite large, while global quantities (like network diameter) essentially take the values of random networks as soon as the fraction if rewiring is greater than zero. What happens to community detection?

\section{The CA network}\label{sec:network}
We shall consider $N$ individuals, labelled from $1$ to $N$.  The nodes are divided in a number $G$ of groups, and each group in a number $C$ of  communities. The nodes in each community are connected using a local connection scheme, like in standard cellular automata of connectivity $K$: node $i$ is connected to nodes $j=i-K/2, i-K/2+1,\dots, i+K/2$, with periodic boundary conditions inside each community. The communities are therefore initially separated. 

With probability $p$, each link can be rewired. Once detached it is reattached to a random node inside the group with probability $q$ of in the whole network with probability $1-q$. In this way we can build networks with a three-level structure: community, groups of communities, whole network.

The network is represented by the adjacency matrix  $A_{ij}=1$ $(0)$ is nodes $i$ takes information from node $j$ (the matrix $A$ needs not to be symmetric).
An example of the adjacency matrix is reported in Figure~\ref{network}. 

We  assume that each individual spends the same amount of time in gathering information, so that people with more connections dedicate less time to each of them. Since the amount of available time is limited, we normalize the adjacency matrix on the index $i$  (i.e., we assign at each link the inverse of the input degree of the incoming node), forming a Markov communication matrix $M$
\begin{equation}
  M_{ij} = \frac{A_{ij}}{\sum_k A_{ik}}.
\end{equation}

\section{The community detection algorithm}
The community detection algorithm is implemented in each node and is updated in parallel, it is therefore a kind of cellular automaton rule. 

Each individual $i$ is characterized by a knowledge vector
a state vector $S^{(i)}$, representing his knowledge of the outer world. 

The vector $S^{(i)}$ is a probability distribution, assuming that $S^{(i)}_j$ is the probability that individual $i$ belongs to the community $j$. Thus, $S^{(i)}_j$ is normalized on the index $j$. We shall denote with $S=S(t)$ the state of the all network at time $t$, with $S_{ij} = S^{(i)}_j$. We shall initialize the system by setting $S_{ij}(0) = \delta_{ij}$, where $\delta$ is the Kronecker delta, $\delta_{ij}=1$ if $i=j$ and zero otherwise. In other words, at time 0 each node knows only about itself.  

The dynamics of the network is given by an alternation of communication and elaboration phases. Communication is implemented as a simple diffusion process, with memory $m$. The memory parameter $m$ allows us to introduce some limitations in human cognitive such as the mechanism of oblivion and the timing effects: in fact the most recent informations have more relevance than informations gained in the past~\cite{Tulving82,Forster84}.

In the communication phase, the state of the system evolves as
\begin{equation}
\label{ref:prima}
  S_{ij}(t+1/2) = m S_{ij}(t) + (1-m) \sum_k M_{ik}S_{kj}(t), 
\end{equation}
where $M$ is the communication matrix. In this phase an individual becomes aware of neighbouring individuals. 

The elaboration  phase is modelled trying to implement a ``take the best'' heuristics~\cite{Gigerenzer1996}. When real people are asked to take a decision, very rarely they weight all  available pieces of information. If there is some aspect that has a higher importance than others, and one item exhibits it, than the decision is taken, otherwise, the second most important factor is considered, etc. In order to implement an adaptive scheme, we exploit a similarity with a competition dynamics among species. 

If two populations $x$ and $y$ are in competition for a given resource, their total abundance is limited. After normalization, we can assume $x+y=1$, i.e., $x$ and $y$ are the frequency of the two species, and $y=1-x$. The reproduction phase is given by $x' = f(x)$, which we assume to be represented by a power $x'=x^\alpha$. For instance, $\alpha=2$ models birth of individuals of a new generation after binary encounters of individuals belonging to the old generation, with non-overlapping generations (eggs laying) \cite{Nicosia09}. 

After normalization
\begin{equation}
 x' = \frac{x^\alpha}{x^\alpha + y^\alpha} = \frac{x^\alpha}{x^\alpha + (1-x)^\alpha}.
\end{equation}
Introducing $z=(1/x)-1$ ($0\le z< \infty$), we get the map
\begin{equation}  
\label{ref:second}
 z(t+1) = z^\alpha(t), 
\end{equation}
whose fixed points (for $\alpha > 1$) are 0 and $\infty$ (stable attractors) and $1$ (unstable), which separates the basins of the two attractors. Thus, the initial value of $x$, $x_0$, determines the asymptotic value, for $0\le x < 1/2$ $x(t\rightarrow\infty) = 0$, and  for $1/2< x < 1$ $x(t\rightarrow\infty) = 1$.

By extending to a larger number of components for a probability distribution $S^{(i)}$, the competition dynamics becomes
\begin{equation}
\label{ref:terza}
  S_{ij}(t+1) = \frac{ S_{ij}(t+1/2)^\alpha}{\sum_j  S_{ij}^{\alpha}(t+1/2)},
\end{equation}
and the iteration of this mapping, for $\alpha>1$,   leads to a Kronecker delta, corresponding to the largest component.  However, the convergence time depends on the relative differences among the components and therefore, when coupled with the information propagation phase, it can originate interesting behaviours. 

The model therefore has two free parameters, the memory $m$ and  the exponent $\alpha$. 

\begin{figure}[t!]
\centering
\subfigure[] 
{\includegraphics[width=6cm]{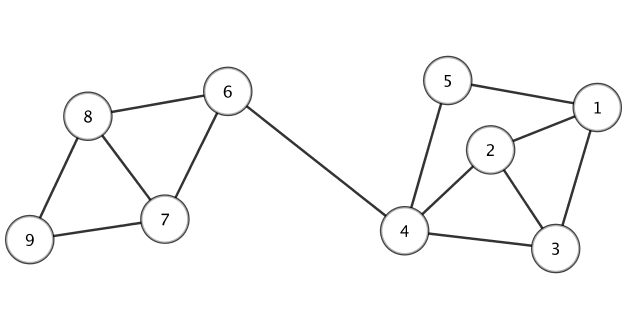}}
\hspace{5mm}
\subfigure[]
{\includegraphics[width=6cm]{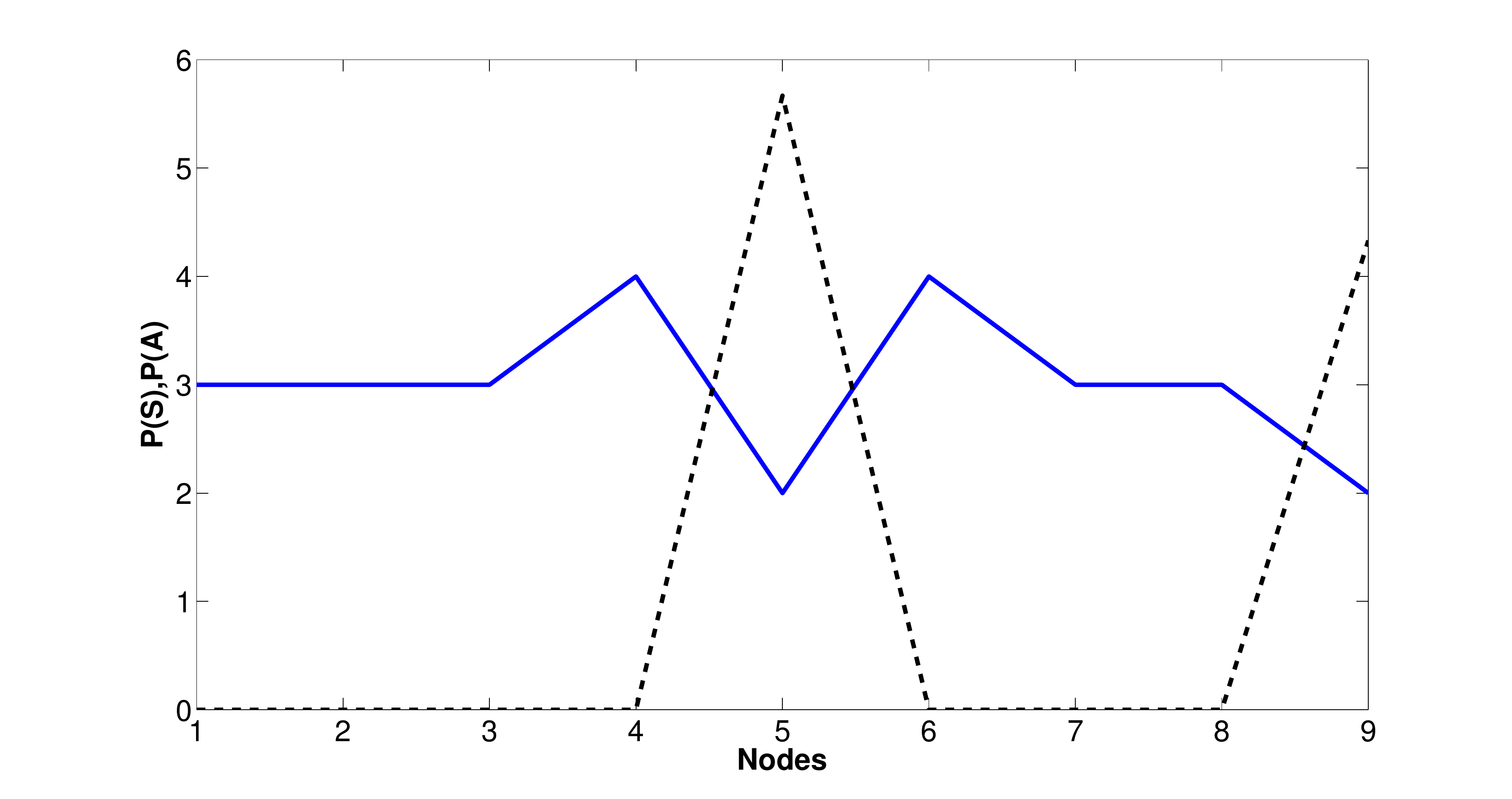}}
\hspace{5mm}
\subfigure[]
{\includegraphics[width=6cm]{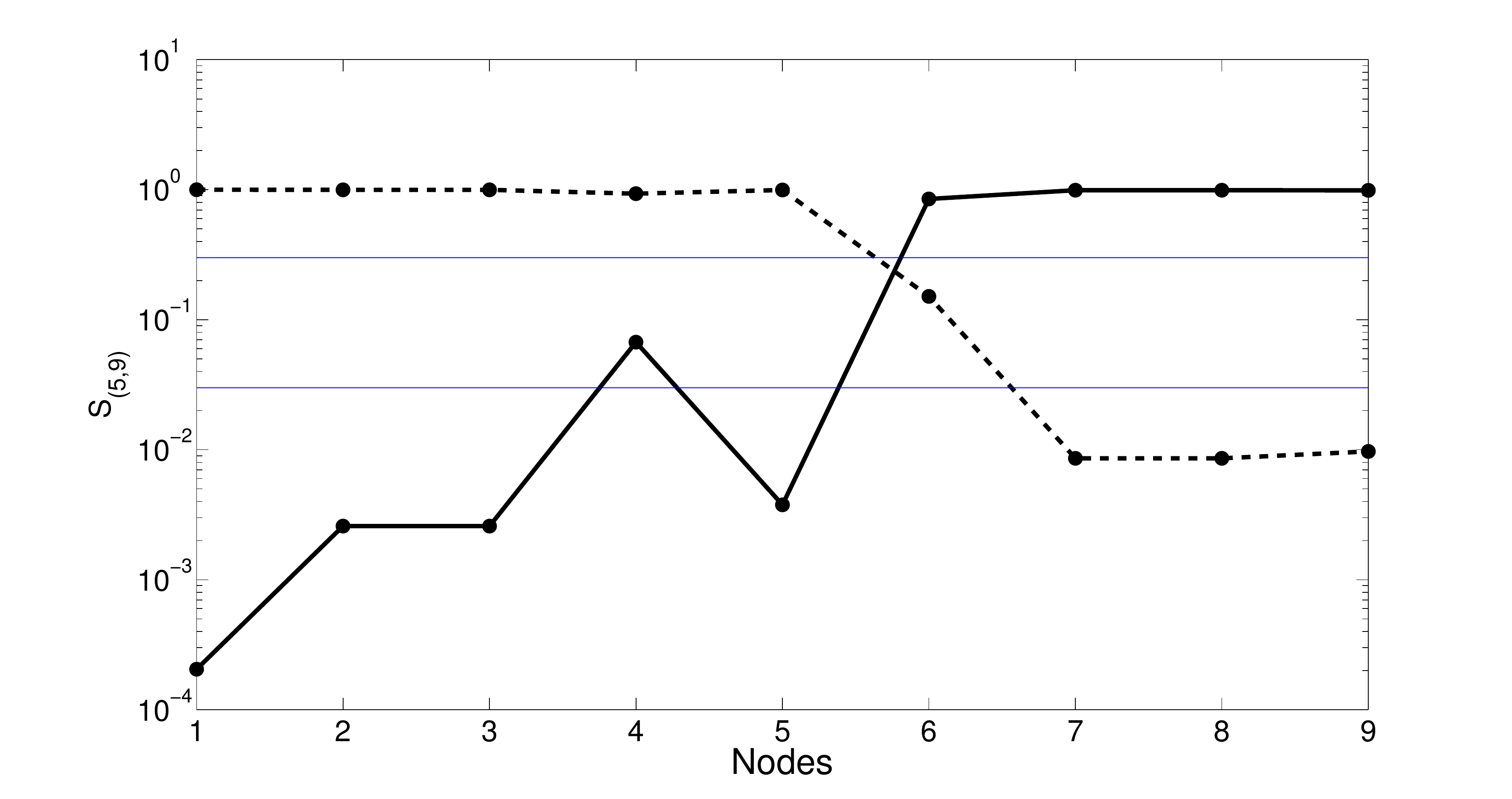}}
\caption{\label{fig:cfart1} (a) \emph{Synthetic network} composed by of 9 nodes and 13 links divided in 2 communities. It is possible to identify two different communities: the first one composed by nodes 1-2-3-4-5 and the second one by 6-7-8-9. (b) On the x-axis of both figures there is the node index. On the y-axis there are the cumulative distributions  $P^{(S)}$ (dashed black line, $P_j^{(S)}=\sum_i S_{ij}$, multiplied by five) and $P^{(A)}$ (grey line, $P_j^{(A)}=\sum_i A_{ij}$, connectivity). The information propagation algorithm identifies communities by leaves (nodes 5 and 9 with lower connectivity) with $m=0.3$ and $\alpha = 1.4$. (c) The value of state vectors, at the final asymptotic time, of node 5 (dashed black line) and node 9 (black line). We can observe upper values identifying communities: the first one composed by nodes 1-2-3-4-5 and the second one by nodes 6-7-8-9. The algorithm is capable also to detect the \emph{communication nodes} 4 and 6 between the grey lines.  In this way we can identify the overlap between the communities and also define a sort of \emph{objective vision} of nodes. It is clear that the upper nodes know very well which is their community as well as nodes 4 and 6 that know that they are in a middle state between two communities.}
\end{figure}

Using a simple synthetic network, Figure~\ref{fig:cfart1}(a), it is possible to explain our algorithm; it faces with a very simple task and converges to an optimal solution in few iterations and for a wide range of model's parameters $m$ and $\alpha$. Analysing the state matrix S(t), it is possible to identify two different communities marked by nodes 5 and 9.

In Figure~\ref{fig:cfart1} (c) it is possible to identify two different communities highlighted by upper values in the graph. The first community is composed by node 1-2-3-4-5 and the second one by 6-7-8-9. Our algorithm is capable also to detect overlapping nodes (4 and 6) as "middle" values between grey lines. In this way each node knows exactly its role in the network.

In order to summarize the amount of information gathered by all nodes, we compute the Shannon entropy of the knowledge of nodes: we build the distribution of labels $P_j$ by summing the knowledge matrix $S_{ij}$ over the index $i$ ($S_{ij}$ is the amount of information of site $i$ about site $j$), 
\[
	P_j = \frac{1}{N}\sum_i S_{ij}
\]
and compute 
\[
  H = -\sum_j P_j \log(P_j). 
\]
In case of a population formed by $n$ communities of the same size, each characterized by a different label (so that only $n$ values $P_j=1/n$ are different from zero), we get $H=\log(n)$. There are therefore four characteristic values of $H$: $\log(N)$ if each node knows only about itself (this is the starting value), $\log(N/(GC))$ is the nodes organize their information at the level of communities, $\log(N/G)$  if the nodes cluster their knowledge at the level of groups, and zero if the nodes share a common label (only one big community).

In order to make the results more easily readable, we plot $\exp(H)$ as a function of time, so that the reference values are now $N$, $N/(GC)$, $N/G$ and $1$. 
\subsection{Results}
We have generated  matrices as defined in Section~\ref{sec:network}, with $K=5$, $N=120$, $G=3$, $C=2$ ($3$ groups of $40$ nodes, and communities of $20$ nodes).  After having generated the networks with uniform local connectivity $K$, links are rewired with probability $p_r$. If rewired, the site is connected to another one (possibly already connected) in the same community with probability $p_c$, in the same group with probability $(1-p_c)p_g$ and to a random node with probability $(1-p_c)(1-p_g)$.  
\begin{figure}[t!]
\centering
\subfigure[] 
{\includegraphics[width=0.35\columnwidth]{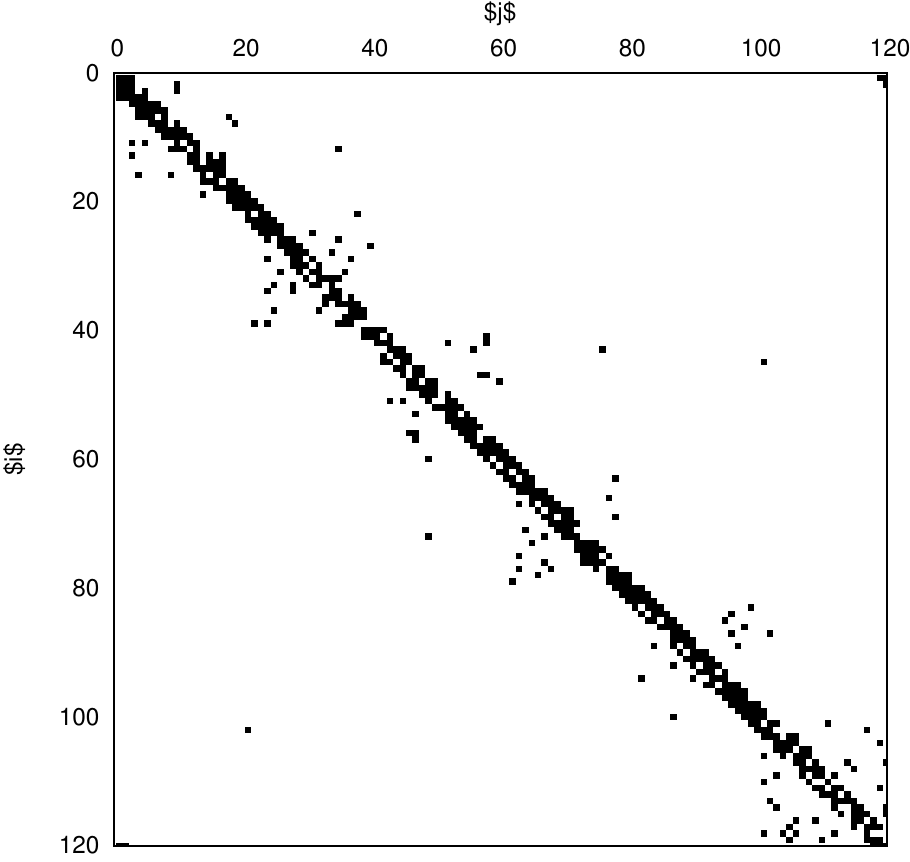}}
\hspace{1mm}
\subfigure[]
{\includegraphics[width=0.6\columnwidth]{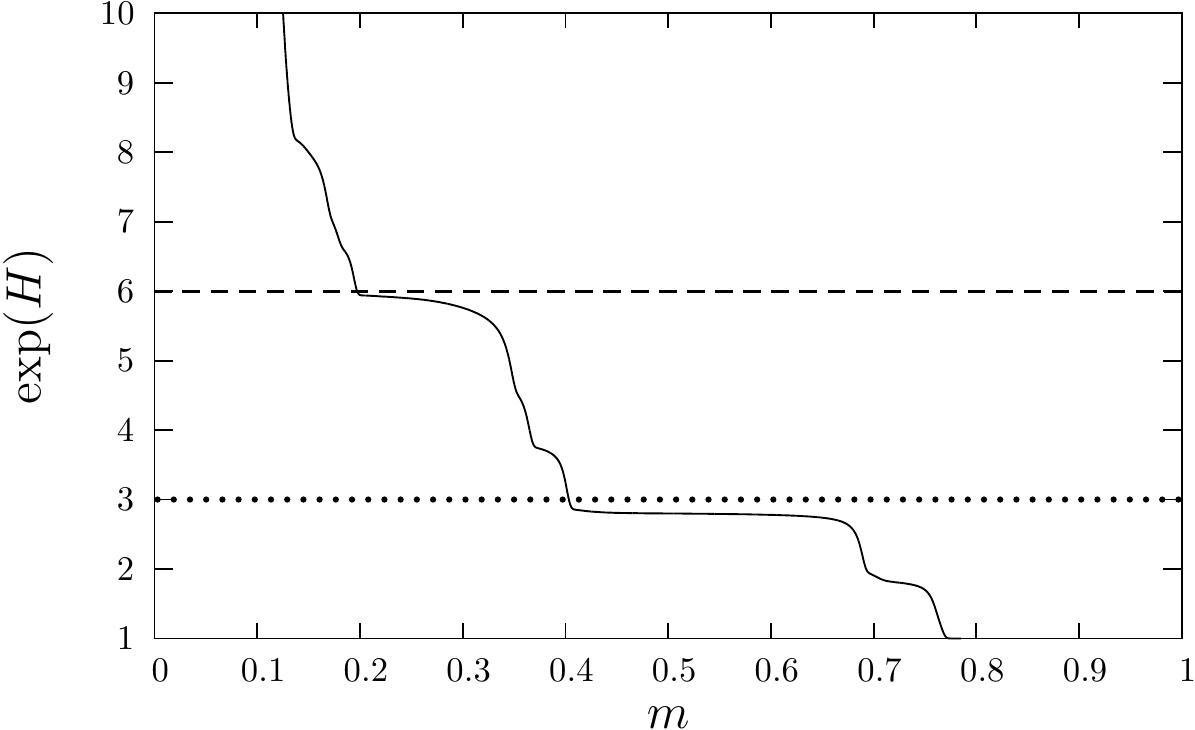}}
\caption{\label{network} (a) One adjacency matrix for $K=5$, $N=120$, $G=3$, $C=2$, $p_r=0.2$, $p_c=0.9$, $p_g=0.7$. Black dots corresponds to $A_{ij}>0$. (b) The plot of $\exp(H)$ vs $m$ for the network of Figure~\ref{network}(a). The dotted line marks the  value of $\exp(H)$ corresponding to the three groups level and the dashed line marks the value corresponding to the six communities level. }
\end{figure}

An example of such matrix is reported in Figure~\ref{network}-a. The rewiring probabilities ($p_r=0.2$, $p_c=0.9$, $p_g=0.7$) are such that the local structure is extremely evident, followed by the community structure. The group structure is almost invisible.

In order to reveal all structures of communities, we have slowly varied $m$, for a given value of $\alpha$. The community structure for $\alpha=1.04$ is reported in Figure~\ref{network}-b. The levels of $\exp(H)$ corresponding to the group and community structures are marked. By changing the value of the memory $m$, nodes tend to accumulate more knowledge about the external world, and, due to the competition phase, their memory becomes dominated in general by just one label, that marking the community the node belongs to. There are occasional transitions when two communities fuse together, in the sense that a label from one community invades the other. It is possible to see these transitions. In particular, plateaus corresponding to a structure in six and three communities are evident for large intervals of $m$.  The final state corresponds to just one community (we have not reported the trivial initial phase, with $H\simeq \log(120)$). 
\begin{figure}[t!]
\centering
\subfigure[] 
{\includegraphics[width=0.35\columnwidth]{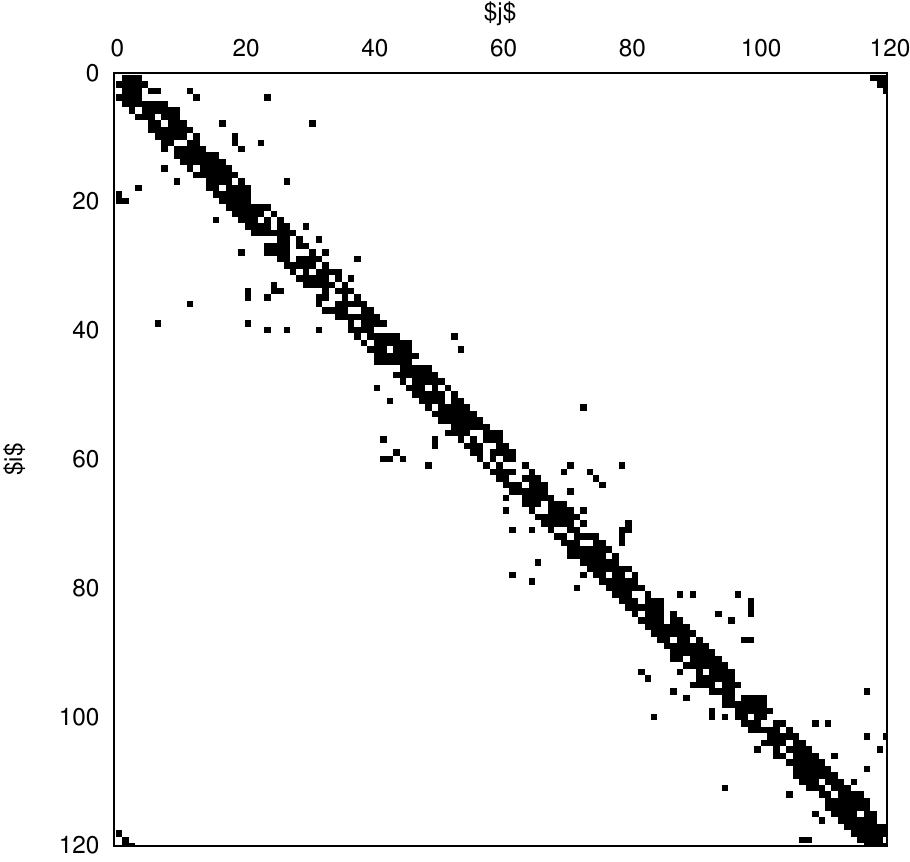}}
\hspace{1mm}
\subfigure[]
{\includegraphics[width=0.6\columnwidth]{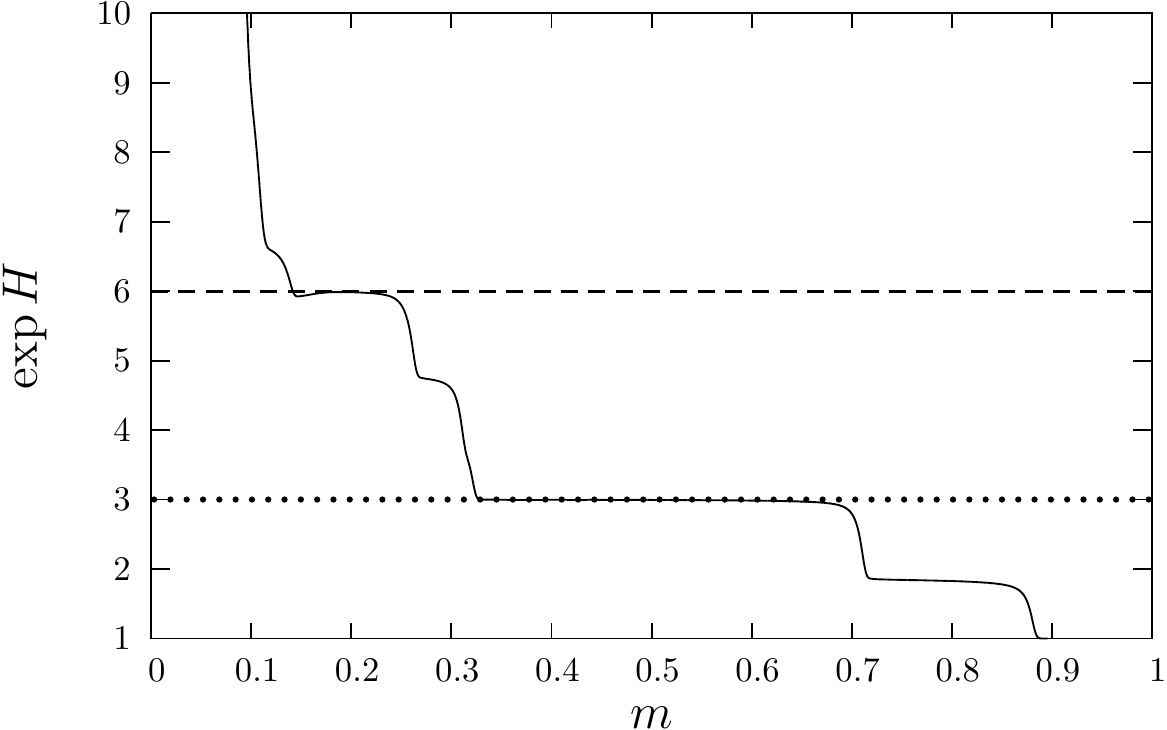}}
\caption{\label{network1} (a) One adjacency matrix for $K=7$, $N=120$, $G=3$, $C=2$, $p_r=0.2$, $p_c=0.9$, $p_g=1.0$. Black dots corresponds to $A_{ij}>0$. (b) The plot of $\exp(H)$ vs $m$ for the network of Figure~\ref{network1}(a). The dotted line marks the  value of $\exp(H)$ corresponding to the three groups level and the dashed line marks the value corresponding to the six communities level. }
\end{figure}

Since the matrices are generated stochastically, it may happen that two communities are more connected in one realization, and therefore the plateaus may happen for slightly different values of $H$. 

Actually, the long-range connections at the network levels are not strictly needed: due to the local connectivity all nodes are connected, and we can set $p_g=1$ and still have the transition to a single community, but this is favored by a larger local connectivity $K$. See for instance the Figure~\ref{network1} . 

\begin{figure}[t!]
\centering
\subfigure[] 
{\includegraphics[width=0.45\columnwidth]{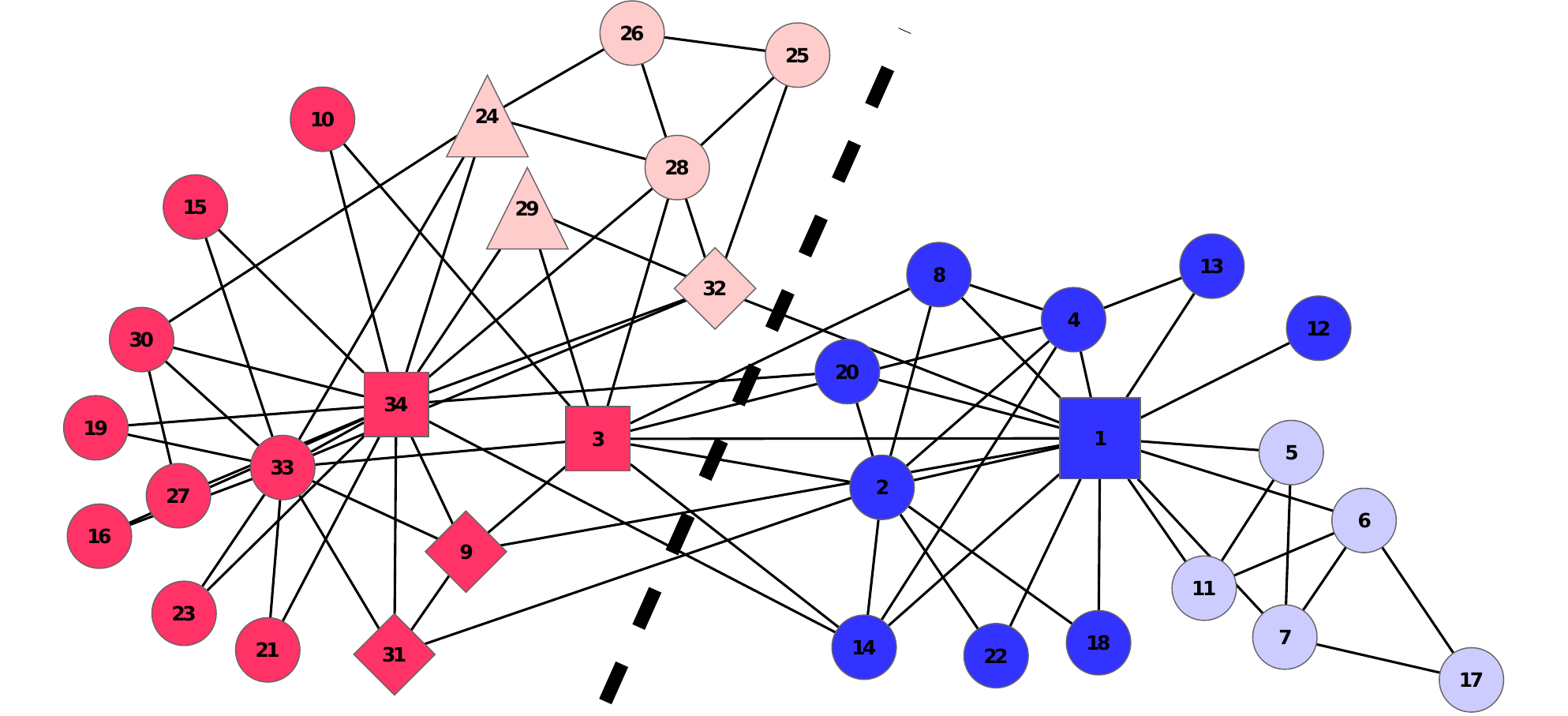}}
\hspace{1mm}
\subfigure[]
{\includegraphics[width=0.45\columnwidth]{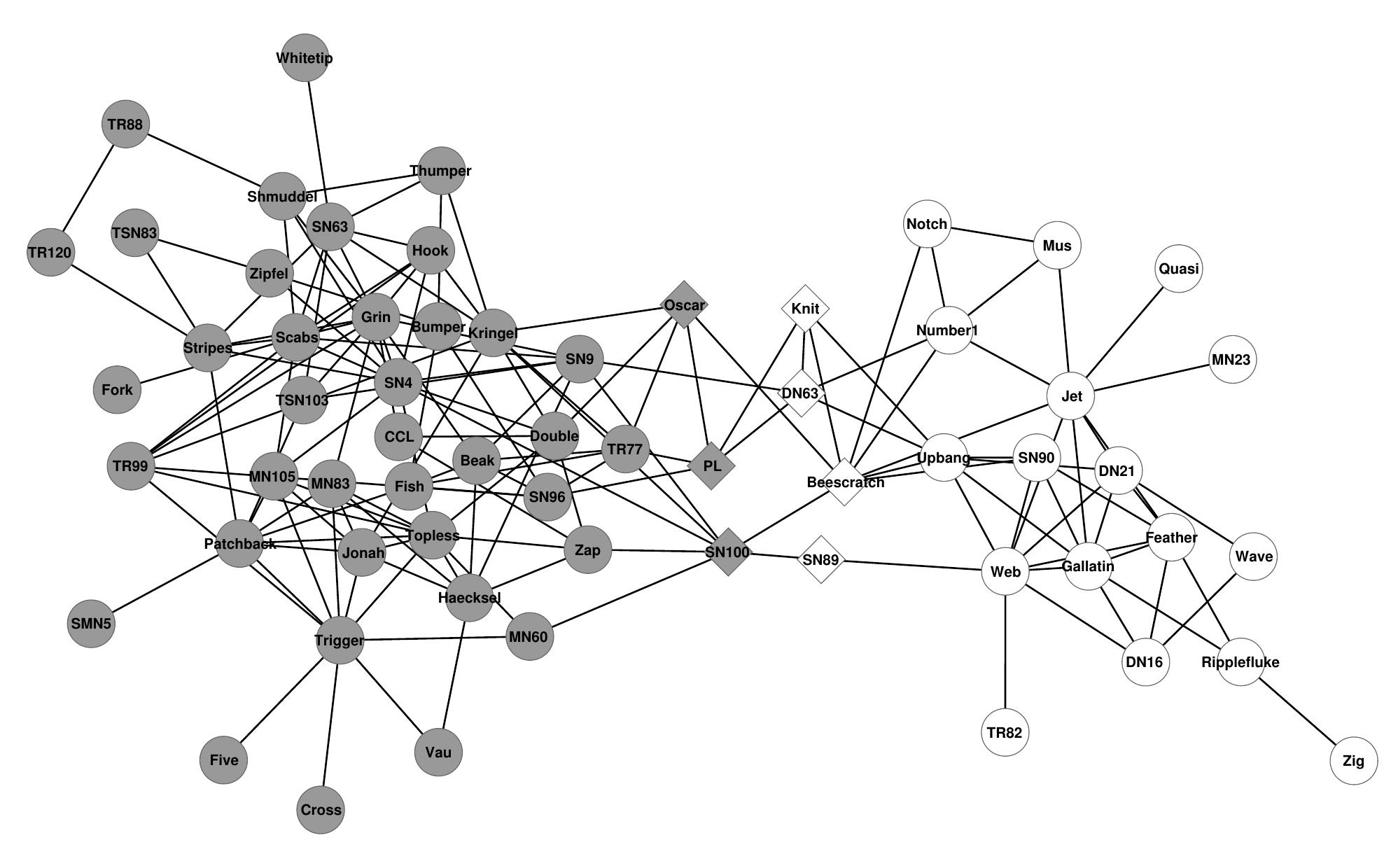}}
\caption{\label{reali} (a) Zachary's karate club network ($m=0.2$ and $\alpha = 1.4$).  (b) Bottlenose dolphin network ($m=0.5$ and $\alpha = 1.03$).}
\end{figure}

We have also applied our method to  two real networks, the  well-known Zachary \emph{karate club} network, Figure~\ref{reali}-a~\cite{Zachary}, and the social interaction of bottlenose dolphins observed by Leusseau, Figure~\ref{reali}-b~\cite{dolphin}. 

For the Zachary club, our algorithm  identifies four communities with different overlapping nodes between them. Considering the hierarchical structure of the network it is possible to merge together two sub-communities. Diamonds denote the overlapping nodes between the two principal communities. Triangles mark the overlapping nodes between the two sub-communities while square are the overlapping nodes between both subcommunities and communities.

The bottlenose dolphin network has a size of 62 nodes and was obtained by direct observation. Our algorithm detects 2 principal communities but also 7 overlapping nodes (diamonds) between them.

\begin{figure}[t!]
\centering
\subfigure[] 
{\includegraphics[width=2.5cm]{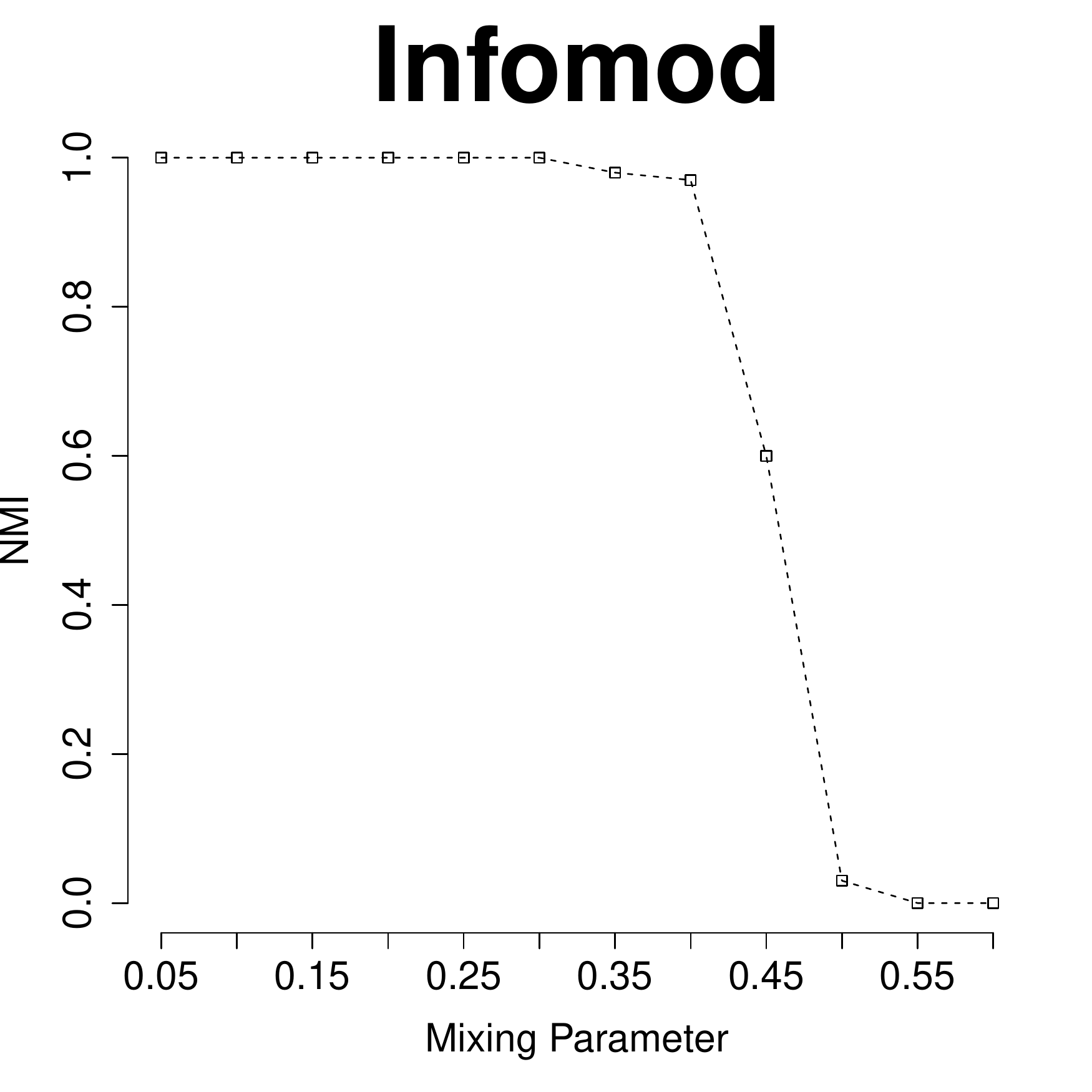}}
\hspace{1mm}
\subfigure[]
{\includegraphics[width=2.5cm]{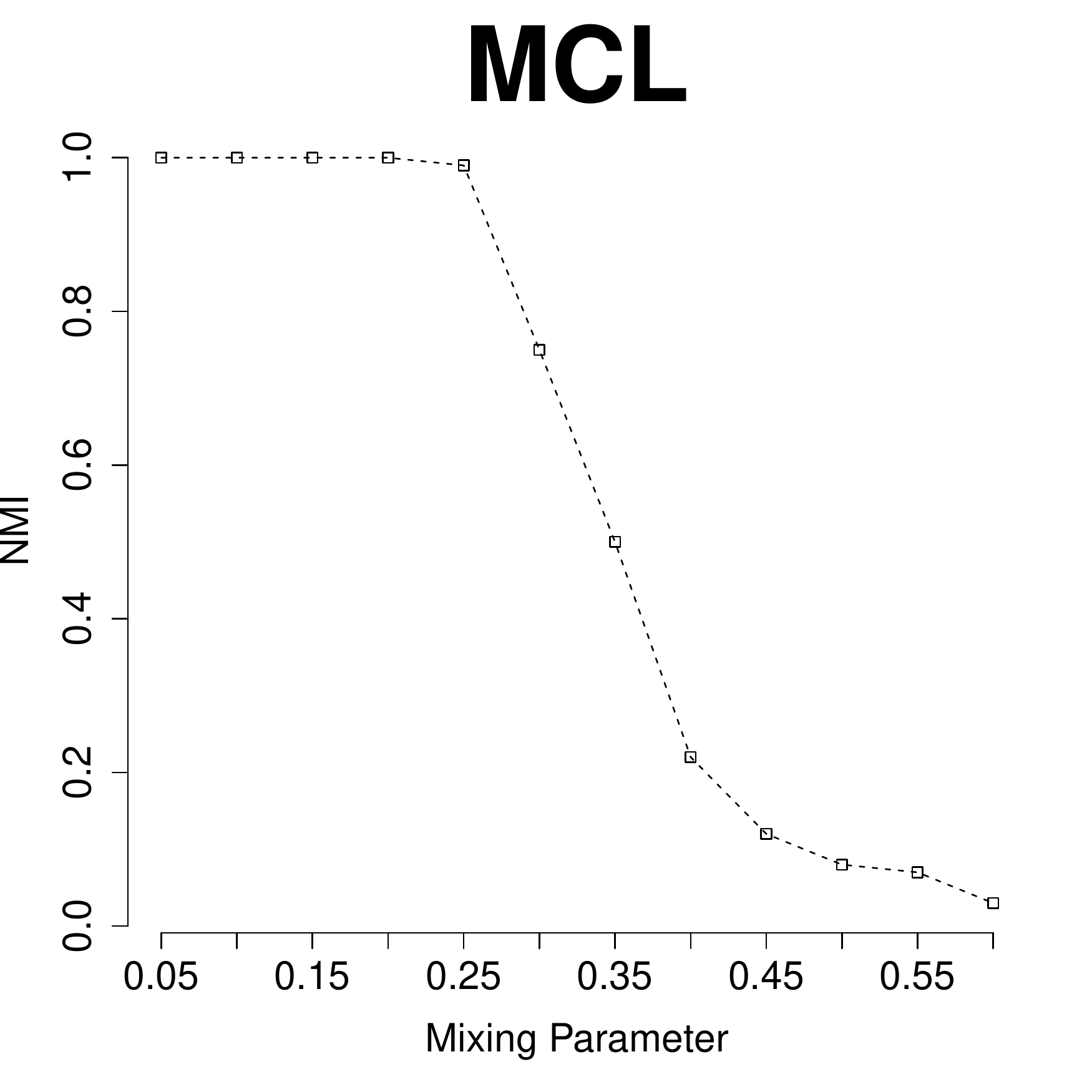}}
\hspace{1mm}
\subfigure[]
{\includegraphics[width=2.5cm]{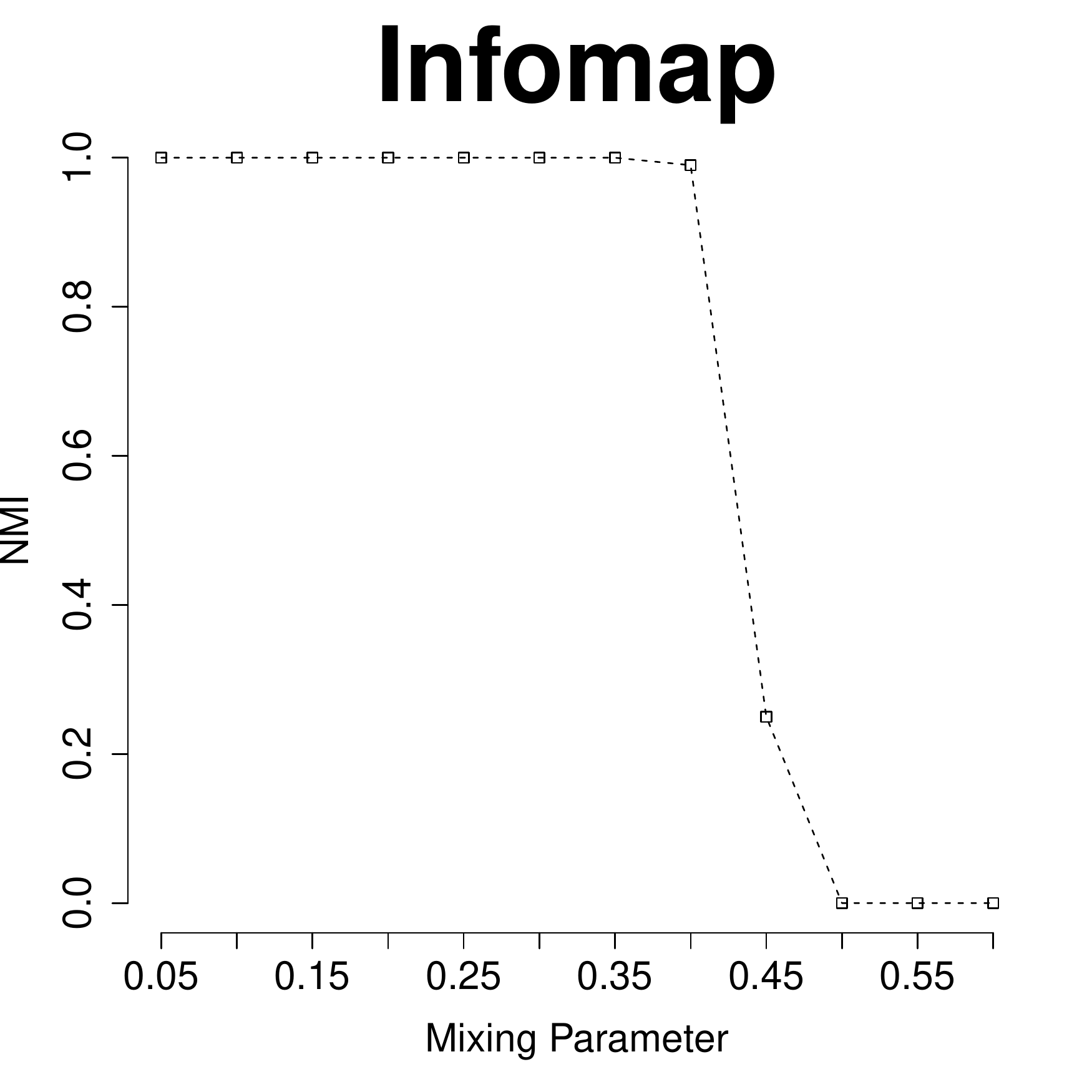}}
\hspace{1mm}
\subfigure[]
{\includegraphics[width=2.5cm]{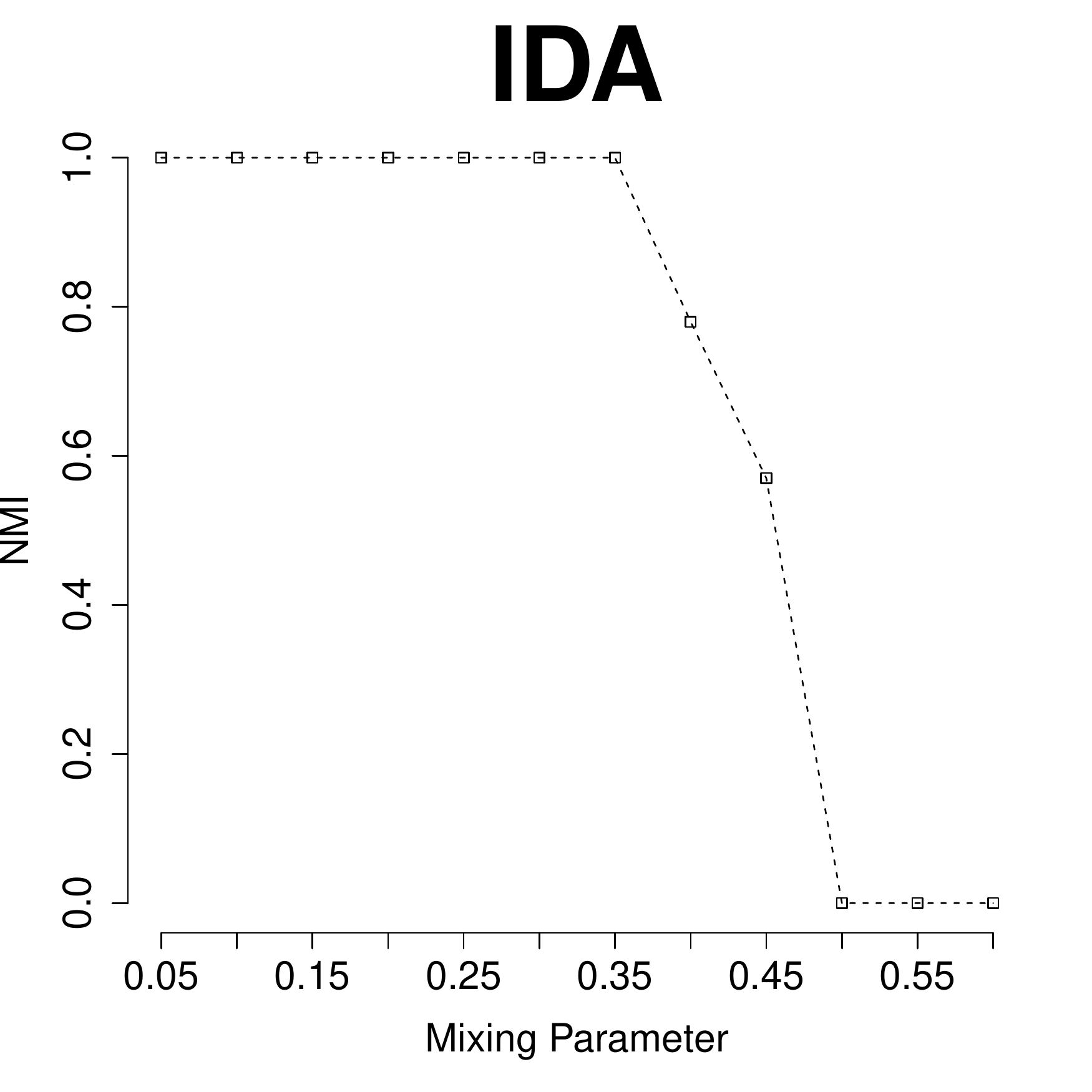}}
\caption{\label{fig:NMI} Test of the algorithm on the GN benchmark based on normalized mutual information (NMI) on the y axes, and the mixing parameter $\mu$ on the x axes.  (a) Infomod~\cite{Infomod}, (b) MCL~\cite{MCL}  (c) Infomap~\cite{Infomap}, (d) our model.}
\end{figure}

Finally, for evaluating our algorithm's performance  we computed the normalized mutual information (NMI) on a Girvan-Newman (GN) benchmark  graph~\cite{Girvan02} varying the mixing parameter $ \mu$. An important benchmark, for testing community detection algorithms, is the model proposed by Girvan and Newman~\cite{Girvan02}, see Figure~\ref{fig:NMI}. Here the graph consists of 128 nodes, each with expected degree 16, which are divided into four groups of 32. The GN benchmark
is regularly used for testing algorithms for community detection. 

We created 11 networks varying the mixing parameters $\mu$, and compared the results with other well knows community-detection algorithms. 
 We performed simulations  with different values of parameters $m$ and $\alpha$. Results (Figure~\ref{fig:NMI}-(d)) show that our algorithm achieves very good performance: in fact, up to $\mu = 0.35$ it always finds the predefined partition in four clusters. In the Figure~\ref{fig:NMI}-(a)-(b)-(c) we reported the results achieved by Lancichinetti and Fortunato~\cite{Lanc_comp} on three well-known community detection algorithms.

A final remark concerns the memory requirement of our cellular automata. 
We have chosen here the simplest implementation by furnishing to all nodes enough memory to contain the whole network (\textit{i.e.}, $S$ is a $N\times N$ matrix), but in practice the number of entries different from zero are always quite few. It is  therefore possible to assume that the nodes have bounded memory, as required by the ``prescriptions'' of human heuristics. 

\section{Conclusions}

We have implemented a community-detection algorithms inspired by human heuristics, as a cellular automaton with some long-range rewiring. In spite of a possible ``small world effect'',  we have seen that it is possible to tune the parameters so to have ``windows'' in which the nodes of the network adopt the label of different communities.

The main advantage of our method is precisely that of furnishing different ``views'' of the clustering levels from an individual point of view, \textit{i.e.}, reveal the structure of nested communities an individual belongs to.

\section*{Acknowledgements}
We acknowledge fruitful discussions with P. Lió and Dr. A. Lancichinetti. This work is financially supported by Recognition Project RECOGNITION, a 7th Framework Programme project funded under the FET initiative.

\bibliography{biblio}{}

\bibliographystyle{plain}
\end{document}